\documentclass[12pt]{article}
\usepackage[english]{babel}
\usepackage{amssymb}
\usepackage{amssymb,amsmath,amsfonts}
\usepackage{mathtools}

\newcommand\btheorem{\vspace{0.31em}\par\refstepcounter{thrm}\bf T\,h\,e\,o\,r\,e\,m\,~\thethrm\,\,~\sl}
\newcommand\etheorem{\rm\vspace{0.3em}\par}
\newcommand\bbtheorem{\vspace{0.31em}\par\refstepcounter{thrm}\bf T\,h\,e\,o\,r\,e\,m\,~\thethrm\,.\,~\sl}
\newcommand\ebtheorem{\rm\vspace{0.3em}\par}
\newcommand\proofr{{\it Proof.\,~}}
\def\proofend{$\Box$\vspace{0.3em}\par}

\title { A generalized concatenation  construction for $q$-ary 1-perfect codes }
\author {Alexander~M.~Romanov\thanks{Sobolev Institute of Mathematics, Russian Academy of Sciences,  4 Academician Koptyug avenue, 630090 Novosibirsk, Russia.  Email: rom@math.nsc.ru}
}
\date{}

\begin{document}

\maketitle

\abstract{ We consider perfect  1-error correcting codes  over a finite field with $q$ elements (briefly  $q$-ary 1-perfect codes). In this paper, a generalized concatenation construction for $q$-ary 1-perfect codes is presented that allows  us  to construct $q$-ary 1-perfect codes of length $(q - 1)nm + n + m$ from the given $q$-ary 1-perfect codes of length  $n =(q^{s_1} - 1) / (q - 1)$ and $m = (q^{s_2} - 1) / (q - 1)$, where $s_1, s_2$ are  natural numbers not less than two. This  construction  allows  us  to also construct $q$-ary codes with
parameters $(q^{s_1 + s_2}, q^{q^{s_1 + s_2} - (s_1 + s_2) - 1}, 3)_q$ and   can be regarded as a $q$-ary analogue of the well-known Phelps construction.}

\section{Introduction}\label{sec:intr}

Let $\mathbb{F}_{q}^n $ be a vector space of dimension $n$ over the finite field $\mathbb{F}_{q}$.
The  \emph{Hamming distance} between two vectors ${\bf x}$, ${\bf y} \in \mathbb{F}_{q}^n$ is equal to the number of coordinates in which they differ and is denoted by $d({\bf x}, {\bf y})$.
An arbitrary subset ${\mathbb C}$ of $\mathbb{F}_{q}^n$ is called a $q$-\emph{ary $1$-perfect code} if for each vector ${\bf x} \in \mathbb{F}_{q}^n$ there exists a unique vector ${\bf c} \in {\mathbb C} $ such that
$d({\bf x},  {\bf c}) \leq 1$.
Nontrivial $q$-ary $1$-perfect codes of length $n$ exist only for
$n = (q^{s} - 1) / (q-1)$, where $s$ is a natural number not less than two.

Two codes ${\mathbb C}_1, {\mathbb C}_2 \subseteq \mathbb {F}_{q}^n$ are called \emph{equivalent} if there exists a vector ${\bf v} \in \mathbb {F}_{q} ^ n $ and a monomial matrix $M$ of size $n \times n$ over the field
$\mathbb{F}_{q}$ such that
$${\mathbb C}_2 =\{({\bf v} + {\bf c}M) \, \,  |  \, \,  {\bf c} \in  {\mathbb C}_1  \}.$$
We assume that the zero vector ${\bf 0}$ belongs to the code. A code is called \emph{linear} if it forms a linear subspace over $\mathbb{F}_{q}$.
Linear $q$-ary $1$-perfect code of length $n$ is unique up to equivalence and is called
$q$-\emph{ary Hamming code}.
Linear $q$-ary $1$-perfect codes of length $n$ exist for all $n =(q^{s} - 1) / (q - 1)$, where $s$ is a natural number not less than two.
Non-linear $q$-ary $1$-perfect codes of length $n = (q^s - 1) / (q - 1)$ exist for $q = 2$, $s \geq 4$; $q \geq 3$,
$s \geq 3$; $q \geq 5$, $s \geq 2$;  see \cite{lin, ph51, ph5, rom7}.
Non-linear $q$-ary $1$-perfect codes of length $n = (q^s - 1) / (q - 1)$  does not exist for $q = 2$, $s \leq 3$;
$q = 3$, $s = 2$; $q = 4$, $s = 2$, see \cite{ald}.

The  \emph{rank} of the code $\mathbb C$ is the maximum number of linearly independent codewords.
It is said that a code of length $n$ and rank $n$ is a  code of  \emph{full rank}; otherwise, the code is a
\emph {non-full-rank} code.

The switching constructions of $q$-ary $1$-perfect codes of full rank are proposed
for all $n = (q^{s}-1)/(q-1)$, where $s \geq 4$, see \cite{etz,  ph5, rom}.
For $s = 3$ and for $q = p^r$, $r > 1$ (where $p$ is a prime number, $r$ is a positive integer) the existence of
$q$-ary $1$-perfect codes of full rank is proved in \cite{ph51}.
The existence of full-rank $q$-ary $1$-perfect codes of  length  $n = (q^s-1) / (q-1)$ still remains open if $s = 3$, $q \geq 3 $, $q$ is a prime number, and if $s =2 $, $q \geq 5$, see \cite {ph51, ph5}.

The switching construction is closely related to the question of  the minimum and maximum possible cardinality of the intersection of two distinct  $1$-perfect codes of the same length.
In the $q$-ary case,  this question still remains open.
In the binary case,  this question was answered in \cite{etz, etz2}.

It is established  that there exist at least $q^{q^{cn}}$  nonequivalent $q$-ary  1-perfect codes of length $n$, where $c = \frac{1}{q} - \epsilon $, see \cite{lin, sch,  vas1}.

Let a $q$-ary code of length $n$ contain $M$ codewords and the minimum distance of this code is $d$. Then the following inequality holds:
\begin{align}
\label{eq:1}
M \leq q^{n -d + 1}.
\end{align}
The inequality (\ref{eq:1}) is called the  \emph{Singleton bound}.
The codes that achieve the Singleton bound are called \emph{maximum distance separable codes} or briefly  MDS.
The  MDS  codes  with parameters $[n,1,n]_q, \, [n, n-1,2]_q, \, [n,n,1]_q$ are called \emph{trivial} MDS codes.
It is widely known that MDS codes are the same as orthogonal arrays of index unity.
In this paper, only codes with a minimum distance of 2 will be considered.

We give a description of the \emph{concatenation construction} of  $1$-perfect codes for the binary case.
Let $s$ be any natural number not less than two and $n = 2^{s} - 1$.
Next, let $\mathbb{C}_{0}^1, \mathbb{C}_{1}^1, \ldots, \mathbb{C}_{n}^1$  be a partition of the vector space $\mathbb{F}_{2}^{n}$ into binary 1-perfect codes of length $n$ and
let $\mathbb{C}_{0}^2, \mathbb{C}_{1}^2, \ldots, \mathbb{C}_{n}^2$ be a partition of the binary MDS code with parameters $[n + 1, n, 2]$ to binary extended 1-perfect
codes with parameters $(n + 1, 2^{n - s}, 4)$.
(In the binary case, the  MDS code with parameters $[n + 1, n, 2]$ consists of all binary vectors of length $n + 1$ of even weight.)

Then the given partitions  $\mathbb{C}_{0}^1, \mathbb{C}_{1}^1, \ldots, \mathbb{C}_{n}^1 $,  \,
$\mathbb{C}_{0}^2, \mathbb{C}_{1}^2, \ldots, \mathbb{C}_{n}^2 $
and the permutation $\alpha$, acting on the set of indices $I = \{0, 1, \ldots , n\}$, defines a binary 1-perfect code

$$\mathbb{C}_{\alpha} = \{({\bf u}| {\bf v}) \, \,  |  \, \, {\bf u} \in \mathbb{C}_{i}^1 , {\bf v} \in \mathbb{C}_{\alpha(i)}^2   \}$$
of length $2n + 1$, where $(\cdot|\cdot)$ denotes concatenation.

In the binary case, the concatenation construction is based on partitions of two types -- a partition of the space $\mathbb {F}_{2}^{n}$ into 1-perfect codes of length $n$ and a partition of the  binary MDS code of length
$n + 1$ into extended 1-perfect codes of length $n + 1$.
In   \cite{rom2}, a direct generalization of the binary concatenation construction to the $q$-ary case is given and the parameters of codes that correspond to binary extended 1-perfect codes in the $q$-ary case are found.
In  the $q$-ary case, codes with parameters $((q - 1)n + 1, q^{(q - 1)n - s}, 3)_q $ correspond to the binary extended 1-perfect codes of length $n + 1$, see \cite{rom2}.
In this paper we give a direct generalization to the $q$-ary case of the well-known Phelps construction  \cite{ph4}.
A review of Phelps's constructions, as well as generalizations, is given in Section~\ref{sec:prel}.
Section~\ref{sec:const}  presents two new constructions.
This is a generalized  concatenation  construction of $q$-ary 1-perfect  codes of length
$(q - 1)nm + n + m$ and a generalized  concatenation construction of $q$-ary codes with
parameters $(q^{s_1 + s_2}, q^{q^{s_1 + s_2} - (s_1 + s_2) - 1}, 3)_q$. The second construction  can be regarded as a $q$-ary analogue of the well-known Phelps construction \cite{ph4}.

\section{Phelps's constructions  and generalizations}\label{sec:prel}

First, we give some definitions.
The most popular methods for constructing non-linear 1-perfect codes are the switching methods
(see, for example, \cite{etz, lin,  sch, vas1}).
In a $q$-ary 1-perfect code of length $n$, a  proper  subset of codewords is allocated and this subset of words is replaced by another subset of words of the same length $n$ over an alphabet consisting of elements of the field $\mathbb {F}_{q}$, so that in the result we get a $q$-ary 1-perfect code of length $n$ that is different from the original one.
The proper subsets of codewords of a $q$-ary  1-perfect  code  that can be replaced  in this way  are called {\it components} of this code.
If the code $\mathbb {C}'$ is obtained from the code $\mathbb {C}$ by replacing a component
of the code $\mathbb {C}$, then we say that the code $\mathbb {C}'$ is obtained from the code $\mathbb {C}$ by using {\it switching}.
A {\it switching class} of a $q$-ary 1-perfect code $\mathbb {C}$  is the set of  (nonequivalent) $q$-ary 1-perfect codes  which contains the code $\mathbb {C}$ and  the set of all (nonequivalent) codes that can be obtained from the code  $\mathbb {C}$  by a sequence of switchings.

Define a function $p : \mathbb{F}_{q}^n \rightarrow \mathbb{F}_{q}$.
If   ${\bf x } = (x_1, x_2, \ldots, x_n)\in \mathbb{F}_{q}^n$,  then
$$p({\bf x}) = \sum_{i = 1}^n x_i.$$
The function $p({\bf x})$ is called {\it parity function}.

Let $\mathbb{C}$ be a $q $-ary code of length $n$.
Then the code
$\widehat{\mathbb{C}} = \{({\bf c} \, | \, p({\bf c})) \, \,  |  \, \, {\bf c} \in \mathbb{C} \}$
has a length equal to $n + 1$ and is called {\it extended code}.
It is said that the extended code $\widehat{\mathbb {C}}$
is obtained from the code $\mathbb {C}$ by adding an overall  parity check.

Now let us consider the concatenation construction.
It is known that the binary extended Hamming codes can be constructed using the well-known
$(\bf u | \bf u + \bf v)$  construction.
The concatenation  construction of binary  1-perfect codes can be considered as a combinatorial generalization of the well-known $(\bf u | \bf u + \bf v)$  construction.
The concatenation construction is closely related to the problem of  the partition of the space
$\mathbb{F}_ {q}^n$ into 1-perfect codes (see, for example, \cite{krot}).
A direct generalization of the binary concatenation construction to the $q$-ary case is obtained in
\cite{rom2}.
It was established in \cite{etz} that in the binary case,   1-perfect codes of full rank can not be constructed using the concatenation construction. In \cite{ph2} with the help of the concatenation construction, binary 1-perfect codes are constructed that belong to different  switching classes. It should be noted that  in \cite{ph2} some specially constructed (not arbitrary) subsets  of the  code C   were considered as components of this code.
Binary 1-perfect codes constructed using concatenation construction were studied by Heden \cite{hed1} and
Solov'eva \cite{sol}.
Phelps proposed a concatenation construction for binary extended 1-perfect codes \cite{ph1}.

In  \cite {ph4} Phelps generalized the concatenation construction for binary extended 1-perfect codes
(Phelps \cite {ph1}), which doubled the length of the code, and instead of permutations suggested using quasigroups; herewith the length of the code began to increase many times.

Let  $n + 1 = 2^{s_1}$.
Consider a binary  MDS code with parameters $[n + 1, n, 2]$ and cosets generated by this code.
Let  $0  \leq k \leq 1$ and let
${\mathbb C}_{0}^k, {\mathbb C}_{1}^k, \ldots, {\mathbb C}_{n}^k$ is a partition of the $k$th coset into binary extended 1-perfect codes with parameters $(m + 1, 2^{m - s_2 }, 4)$.
Further, let $m + 1 = 2^{s_2}$.
Consider the binary extended 1-perfect code $\mathbb R$ with parameters $(m + 1, 2^{m - s_2 }, 4)$.
For each code word ${\bf r} \in \mathbb R$, we define an $m$-ary quasigroup $q_{\bf r}$ of order $n + 1$.
A quasigroup is defined on the index set $\{ 0, 1, \ldots,   n \}$.
Then we form the binary code ${\mathbb C}{\otimes}_q{\mathbb R}$  of length $(n + 1)(m + 1) =  2^{s_1 + s_2}$  as follows:

\begin{multline*}
{\mathbb C}{\otimes}_q{\mathbb R} = \{({\bf c}_0| {\bf c}_1 | {\bf c}_2 | \cdots | {\bf c}_{m} )\, \,  |  \, \,
{\bf c}_i \in {\mathbb C}_{j_i}^{r_i}, {\bf r} = (r_0, r_1, \ldots, r_{m}) \in {\mathbb R}, \\
j_0 =q_{\bf r}(j_1, j_2, \ldots, j_{m}), \,    j_i \in \{ 0, 1, \ldots,   n \},  \,
i = 0, 1, \ldots, m \}.
\end{multline*}

\btheorem {\upshape (Phelps  \cite{ph4})\,\bf{.}}
\label{th1}
The code ${\mathbb C}{\otimes}_q{\mathbb R}$  constructed above is a binary extended 1-perfect code with parameters
$(2^{s_1 + s_2}, 2^{2^{s_1 + s_2} - (s_1 + s_2) - 1}, 4)$.
\etheorem

Historically, the first construction for non-linear binary 1-perfect codes is Vasil'ev's construction \cite {vas1}.

\btheorem {\upshape (Vasil'ev \cite{vas1})\,\bf{.}}
\label{th2}
Let $\mathbb{C}$ be a binary 1-perfect code of length $n$ and let $\lambda$ be a Boolean function defined on the
set $\mathbb {C}$.
Then the set
$$\mathbb{C}_N = \{({\bf u}|{\bf u} + {\bf v}| p({\bf u}) + \lambda({\bf v})  )
\, \,  |  \, \,  {\bf u } \in \mathbb{F}_{2}^n, {\bf v } \in \mathbb{C}  \}$$
is a binary 1-perfect code of length  $N = 2n + 1$.
\etheorem

We observe that the Vasil'ev construction   for $\lambda \equiv 0$ represents  a certain  modification of the
well-known $(\bf u | \bf u + \bf v)$ construction.

Consider the set
$\mathbb{K} = \{ ({\bf u}|{\bf u} | p({\bf u}) )  \, \,  |  \, \,  {\bf u } \in \mathbb{F}_{2}^n \}$.
It's obvious that $\mathbb{K} \subset \mathbb{C}_N$ (since it is assumed that the zero vector ${\bf 0}$ always belongs to the code, then $\lambda ({\bf 0}) = 0 $) and $\mathbb{K}$ is a subspace.
Cosets formed by the subspace $\mathbb{K}$ and belonging to the code $\mathbb{C}_N$ are components of the code  $\mathbb{C}_N$ and form its partition.

Let
${\mathbb K}_{\bf v}  = \{ ({\bf u}|{\bf u} + {\bf v} | p({\bf u}) + \lambda({\bf v}) )
\, \,  |  \, \,  {\bf u } \in \mathbb{F}_{2}^n \}$, ${\bf v } \in \mathbb{C}$.
Then ${\mathbb K}_{\bf v}$  is a component of the code $\mathbb {C}_N$ and the code $\mathbb {C}_N$ is representable as
\begin{equation*}
\mathbb{C}_N =  \bigcup_{{\bf v } \in \mathbb{C}}{\mathbb K}_{\bf v}.
\end{equation*}

A direct generalization of the  Vasil'ev construction to the $q$-ary case  was proposed by Lindstr\"{o}m
\cite{lin} and    Sch\"{o}nheim \cite{sch}.

\btheorem {\upshape (Lindstr\"{o}m \cite{lin}, Sch\"{o}nheim   \cite{sch})\,\bf{.}}
\label{th3}
Let be given  a $q$-ary 1-perfect code  ${\mathbb C}$ of length $n$  and   a function  $\lambda$  defined on the set
${\mathbb C}$ with values in $\mathbb{F}_{q}$.
Let  $\alpha_1, \alpha_2, \ldots, \alpha_{q - 1}$ be all non-zero elements of the field  $\mathbb{F}_{q}$. Then
\begin{multline*}
{\mathbb C}_N =  \Biggl\{\Biggl ({\bf u}_1 | {\bf u}_2 | \cdots | {\bf u}_{q-1} |{\bf v }
+  \sum_{i =1}^{q-1} {\bf u}_i | \sum_{i =1}^{q-1}\alpha_i  p({\bf u}_i) + \lambda({\bf v}) \Biggr) \,\, {\big |} \\
{\bf u}_i \in   \mathbb{F}_{q}^n,   i \in \{1,  \ldots, q-1\},  {\bf v} \in {\mathbb C}  \Biggr\}
\end{multline*}
is a $q$-ary 1-perfect  code of length $N = qn + 1$.
\etheorem
A  generalization of Lindstr\"{o}m-Sch\"{o}nheim's construction is proposed in \cite{rom}.

Further, we move on to  a description of the  construction of Mollard \cite{mol},
which is a generalization of the constructions of Vasil'ev \cite{vas1} and  Phelps \cite{ph4} for binary 1-perfect codes and also is a generalization of the constructions of Lindstr\"{o}m \cite{lin} and Sch\"{o}nheim \cite{sch} for $q$-ary 1-perfect codes.

Let  $\alpha_1, \alpha_2, \ldots, \alpha_{q - 1}$ be all non-zero elements of the field  $\mathbb{F}_{q}$.
Consider a vector ${\bf x} \in \mathbb{F}_{q}^{(q - 1)nm}$.
Let each component of the vector ${\bf x}$ has three  indices $i, j, k$, where
$$i \in \{1, 2, \ldots, q -1\}, \,  j \in \{1, 2, \ldots, n\}, \, k \in \{1, 2, \ldots, m\}.$$

Let the vectors in $\mathbb{F}_{q}^{(q - 1)nm}$ be ordered in lexicographic order according to the indices of their components.
Let $P_1({\bf x})$ be a function defined on the set $\mathbb{F}_{q}^{(q - 1)nm}$ with values in $\mathbb{F}_q^n$.
The function $P_1({\bf x})$ is defined as follows:
$$P_1({\bf x}) = (y_1, y_2, \ldots, y_j, \ldots, y_n),$$
where
$$y_j = \sum_{i = 1}^{q - 1} \sum_{k = 1}^m x_{ijk}.$$
Next, let $P_2({\bf x})$ be a function defined on the set $\mathbb{F}_{q}^{(q - 1)nm}$ with values in $\mathbb{F}_q^m$.
The function $P_2({\bf x})$ is defined as follows:
$$P_2({\bf x}) = (y_1, y_2, \ldots, y_k, \ldots, y_m),$$
where
$$y_k = \sum_{i = 1}^{q - 1} \alpha_i \sum_{j = 1}^n x_{ijk}.$$

\btheorem {\upshape (Mollard   \cite{mol})\,\bf{.}}
\label{th4}
Given a $q$-ary 1-perfect code ${\mathbb C}$ of length $n$,  a $q$-ary 1-perfect code ${\mathbb C}'$ of length $m$, and a vector function $f$, defined on the set ${\mathbb C}$ with values in $\mathbb{F}_{q}^m$, define   the $q$-ary 1-perfect code
$${\mathbb F} = \{ ({\bf x}| {\bf c} + P_1({\bf x})| {\bf c}' + P_2({\bf x}) + f({\bf c})) \, \,  |  \, \,
{\bf x} \in \mathbb{F}_{q}^{(q - 1)nm},  {\bf c} \in \mathbb{C}, \, {\bf c}' \in \mathbb{C}' \}$$
of length $(q-1)nm + n + m$.
\etheorem

Next, we present the Phelps construction \cite{ph6}, which is analogous to the Mollard construction \cite{mol}, but is defined using a series of quasigroups.

Let ${\mathbb Q} = \{({\bf x}|f({\bf x})|g({\bf x})) \, \,  |  \, \, {\bf x} \in \mathbb{F}_{q}^{(q - 1)} \}$
be a $q$-ary 1-perfect code of length $q + 1$. Then $f({\bf x})$ and  $g({\bf x})$ are $(q - 1)$-ary quasigroups of order   $q$.
Let $ F $ be an $(m +1)$-ary quasigroup of order $q$  and $G$ be an $(n + 1)$-ary quasigroup of order $q$.
Let ${\mathbb C}$ and ${\mathbb R}$ be $q$-ary 1-perfect codes of length $n$ and $m$, respectively.
Then we form the code ${\mathbb C}{\otimes}{\mathbb R}$ of length $(q-1)nm + n + m$ as follows:

\begin{multline*}
{\mathbb C}\otimes{\mathbb R} = \{ ({\bf x}_{11}| \cdots |{\bf x}_{ij}| \cdots |{\bf x}_{nm}| F_1|F_2|\cdots|F_n|
G_1|G_2|\cdots|G_m) \, \,  |  \, \, \\
{\bf x}_{ij} \in \mathbb{F}_{q}^{(q - 1)},  {\bf r} \in \mathbb{R}, \,  {\bf c} \in \mathbb{C} \},
\end{multline*}
where \\
$F_i = F(f({\bf x}_{i1}), f({\bf x}_{i2}), \ldots, f({\bf x}_{im}),c_i), \, \, i = 1, 2, \ldots, n,$ \\
$G_i = G(g({\bf x}_{1j}), g({\bf x}_{2j}), \ldots, g({\bf x}_{nj}),r_j), \, \, j = 1, 2, \ldots, m$ \\
and \\
$ {\bf r} = (r_1, r_2, \ldots, r_m),  \, \, i = 1, 2, \ldots, n, $
$ {\bf c} = (c_1, c_2, \ldots, c_n),  \, \, j = 1, 2, \ldots, m. $ \\
The length of the code ${\mathbb C}\otimes{\mathbb R}$   is $(q-1)nm + n + m$.

\btheorem {\upshape (Phelps  \cite{ph6})\,\bf{.}}
\label{th5}
The code ${\mathbb C}{\otimes}{\mathbb R}$ constructed above is a $q$-ary 1-perfect code of length $(q-1)nm + n + m$.
\etheorem

A further generalization of the Phelps construction \cite{ph4} was proposed by Heden and Krotov \cite{hed2}.
Let $n, t, n_1, \ldots, n_t$ be positive integers satisfying the inequality $n_1 + \cdots + n_t \leq n$.
Furthermore, let ${\bf x} = ({\bf x}_1|{\bf x}_2| \cdots |{\bf x}_t|{\bf x}_0) = ({\bf x}_{*}|{\bf x}_0)$,
where ${\bf x}_i = (x_{i1}, x_{i2}, \ldots, x_{in_i})$, $i = 0, 1, \ldots, t$ è $n_0 = n - n_1 - \ldots - n_t$.
Let $$ {\boldsymbol \sigma}({\bf x}) = (\sigma_1({\bf x}_1), \sigma_2({\bf x}_2), \ldots, \sigma_t({\bf x}_t )),$$
where
$$\sigma_i({\bf x}_i) = \sum_{j = 1}^{n_i}x_{ij}.$$
We denote the rank of the code ${\mathbb C}$ by  $\mbox{rank}({\mathbb C})$.
Let  ${\mathbb C}$ be  $q$-ary 1-perfect code of non-full rank that has length $n =  (q^s - 1)/ (q - 1)$.
In \cite{hed2}  it is shown that for any integer $r$ satisfying inequality
$$1 \leq r \leq n - \mbox{rank}({\mathbb C}) $$
there exists a $q$-ary Hamming code   ${\mathbb C}^{\star}$  of length  $t =  (q^r - 1)/ (q - 1)$ such that
for some monomial transformation $\psi$
$$\psi({\mathbb C}) = \bigcup_{{\boldsymbol \mu} \in {\mathbb C}^{\star}} {\mathbb K}_{\boldsymbol \mu},$$
where
$${\mathbb K}_{\boldsymbol \mu} = \{ ({\bf x}_1| {\bf x}_2| \cdots {\bf x}_t|{\bf x}_0) \, \,  |  \, \,
{\boldsymbol \sigma}({\bf x}) = {\boldsymbol \mu}, \, \, {\bf x}_1, {\bf x}_2, \ldots, {\bf x}_t \in  \mathbb{F}_{q}^{q^{s'}}, {\bf x}_0 \in {\mathbb C}_{\boldsymbol \mu}{({\bf x}_{*})}  \}$$
for some family of perfect codes ${\mathbb C}_{\boldsymbol \mu}{\bf(x)}$
of length $(q^{s'} - 1)/ (q - 1)$, where $s' = s - r$, which for each ${{\boldsymbol \mu} \in {\mathbb C}^{\star}}$
satisfy the condition
$$ d({\bf x}_{*},{\bf y}_{*}) \leq 2  \, \,    \, \, \Longrightarrow \, \,    \, \, {\mathbb C}_{\boldsymbol \mu}
({\bf x}_{*})\cap {\mathbb C}_{\boldsymbol \mu}{({\bf y}_{*})} = {\bf 0}.$$
(For $r = s$, the code ${\mathbb C}_{\boldsymbol \mu}$ is a code of length $0$ and cardinality $1$.)
The authors of the paper  \cite{hed2} subset  ${\mathbb K}_{\boldsymbol \mu}$ are called
${\boldsymbol \mu}$-\emph{component} of the code $\psi({\mathbb C})$.

Let $s$ and $r$ be integers, $s > r$; let $n =  (q^s - 1)/ (q - 1)$ and $t =  (q^r - 1)/ (q - 1)$.
Assume that ${{\mathbb C}^{\star}}$ is  a $q$-ary 1-perfect code of length $t$ and for every
${{{\boldsymbol \mu} \in {\mathbb C}^{\star}}}$ we have a  distance-3 code
${\mathbb K}_{\boldsymbol \mu} \subset \mathbb{F}_{q}^{n}$  of cardinality $q^{n - s - (t - r)}$
that  satisfies the following generalized parity-check law:
$$ {\boldsymbol \sigma}({\bf x}) =
(\sigma_1(x_1, \ldots, x_l ),  \ldots, \sigma_t(x_{lt - l + 1} \ldots, x_{lt} )) = {\boldsymbol \mu}$$
for every ${\bf x} = (x_1, \ldots, x_n) \in {\mathbb K}_{\boldsymbol \mu}$,
where $l = q^{s - r}$ and ${\boldsymbol \sigma} = (\sigma_1,  \ldots, \sigma_t)$ is a collections of $l$-ary
quasigroups of order $q$.
Then the union
$${\mathbb C} = \bigcup_{{{\boldsymbol \mu} \in {\mathbb C}^{\star}}}{\mathbb K}_{\boldsymbol \mu}$$
is  a $q$-ary 1-perfect code of length $n$, see \cite{hed2}.

Further, the elements $\sigma_i$ of  ${\boldsymbol \sigma}$ we will represent as the compositions
$V_i(v(\cdot), \ldots, v(\cdot), \cdot)$, where $v$ is a $(q-1)$-ary quasigroup and
$V_i$ is $(k + 1)$-ary quasigroups for some integer $k$.

\btheorem {\upshape (Heden-Krotov \cite{hed2})\,\bf{.}}
\label{th6}
Let  ${\boldsymbol \mu} \in {\mathbb F}_q^t$. Let for every $i$ from 1 to $t + 1$  1-perfect codes
$\mathbb{C}_{i,j}$, $j = 0, 1, \ldots, (q - 1)k$, form a partition of the space ${\mathbb F}_q^k$ and
$\gamma_i : {\mathbb F}_q^k \rightarrow \{0, 1, \ldots, (q - 1)k\}$
be the corresponding partition function:
$$\gamma_i({\bf y}) = j \Longleftrightarrow {\bf y} \in \mathbb{C}_{i,j}.$$
Let $\{({\bf y}|v({\bf y})|h({\bf y})) \, \,  |  \, \, {\bf y} \in \mathbb{F}_{q}^{q - 1} \}$
be a $q$-ary 1-perfect code of length  $q + 1$, where  $v({\bf y})$ and $h({\bf y})$ are $(q - 1)$-ary
quasigroups of order $q$.
Let $V_1, \ldots, V_t$ be $(k + 1)$-ary quasigroups of order $q$ and $G$ be a $t$-ary quasigroup of order
$(q - 1)k + 1$. Then the set
\begin{multline*}
{\mathbb K}_{\boldsymbol \mu} = \{({\bf x}_{11}| \cdots |{\bf x}_{1k}|{y}_{1}|
{\bf x}_{21}| \cdots |{\bf x}_{2k}|{y}_{2}| \cdots |
{\bf x}_{t1}| \cdots |{\bf x}_{tk}|{y}_{t}|
z_1|z_2| \cdots |z_k) \, \,  |  \, \,
\\  {\bf x}_{ij} \in {\mathbb F}_q^{q - 1}, (V_1(v({\bf x}_{11}), \ldots, v({\bf x}_{1k}), y_1), \ldots,
V_t(v({\bf x}_{t1}), \ldots, v({\bf x}_{tk}), y_t)) = {\boldsymbol \mu},
\\G(\gamma_1(h({\bf x}_{11}), \ldots, h({\bf x}_{1k}), \ldots,
\gamma_t(h({\bf x}_{t1}), \ldots, h({\bf x}_{tk}))) = \gamma_{t + 1}(z_1, \ldots, z_k)  \}
\end{multline*}
is a ${\boldsymbol \mu}$-component that satisfies the generalized parity-check law with
$$\sigma_i(\cdot, \ldots, \cdot, \cdot) = V_i(v(\cdot), \ldots, v(\cdot), \cdot).$$
\etheorem
In contrast to the constructions proposed in this paper, the construction of Heden and Krotov  \cite{hed2} is based on partitions of the space ${\mathbb F}_q^k$  into $q$-ary 1-perfect codes and codes with parameters
$((q - 1)n + 1, q^{(q - 1)n - s}, 3)_q $ are not used in it.

\section{Constructions for $q$-ary codes}\label{sec:const}

In this section, we  present a generalized  concatenation construction for $q$-ary 1-perfect codes of
length  $(q - 1)nm + n + m$ and also we   present a generalized concatenation construction for $q$-ary codes with parameters  $(q^{s_1 + s_2}, q^{q^{s_1 + s_2} - (s_1 + s_2) - 1}, 3)_q $.
The second construction can be regarded as a $q$-ary analogue of the well-known Phelps construction \cite{ph4}.

Let  ${\mathbb C}_{0}, {\mathbb C}_{1}, \ldots, {\mathbb C}_{(q - 1)n}$ be a partition of the vector space $\mathbb{F}_{q}^{n}$ into $q$-ary 1-perfect codes of length $n = (q^{s_1}-1)/(q-1)$.
Consider a $q$-ary  MDS code with parameters $[(q - 1)n + 1, (q - 1)n, 2]_q$ and cosets generated by this code.
Let  $0  \leq k \leq q - 1$  and let  $ {\mathbb C}_{0}^k, {\mathbb C}_{1}^k, \ldots, {\mathbb C}_{(q - 1)n}^k$
be a partition of the $k$th coset into $q$-ary codes with parameters
$((q - 1)n + 1, q^{(q - 1)n - s_1}, 3)_q $.
Such  partitions exist \cite{rom2}.

Let $\mathbb R$ be a $q$-ary 1-perfect code of length $m = (q^{s_2}-1)/(q-1)$.
For each code word ${\bf r} \in \mathbb R$, we define an $m$-ary quasigroup $q_{\bf r}$ of order $(q - 1)n + 1$.
A quasigroup is defined on the  index set  $\{ 0, 1, \ldots,   (q - 1)n \}$.
Let $\alpha$ be a primitive element of a finite field  $\mathbb{F}_{q}$.
Consider the $i$th component $r_i$ of the code word ${\bf r} = (r_1, r_2, \ldots, r_m) \in \mathbb R$.
By definition, $r_i \in \mathbb{F}_{q}$.
If $r_i \neq 0$, then $r_i = {\alpha}^k$, where $k \in \{1, \ldots,  q - 1 \}$.
Then the notation ${\mathbb C}_j^{r_i}$ should be understood as ${\mathbb C}_j^{k}$ wherein  $0 \leq j \leq (q - 1)n$. If  $r_i = 0$, then ${\mathbb C}_j^{r_i} = {\mathbb C}_j^0$  and ${\mathbb C}_j^{r_i}$ is an element of the partition of the $0$th coset.

Then we form the $q$-ary code ${\mathbb D}$ of length $(q - 1)nm + n + m$ as follows:

\begin{multline*}
\mathbb D = \{({\bf u}| {\bf v}_1 | {\bf v}_2 | \cdots | {\bf v}_m )\, \, {|}  \, \, {\bf u} \in  {\mathbb C}_{j_0} , {\bf v}_i \in {\mathbb C}_{j_i}^{r_i}   \, \, \mbox{ïðè}   \, \,  1 \leq i \leq m ,
{\bf r} = (r_1, r_2, \ldots, r_m) \in \mathbb R, \\
j_0 =q_{\bf r}(j_1, j_2, \ldots, j_m),  \, \, j_i \in \{ 0, 1, \ldots,   (q - 1)n \} \, \, \mbox{ïðè}   \, \,
i = 0, 1, \ldots, m    \}.
\end{multline*}

\bbtheorem
\label{th7}
The code $\mathbb D$ constructed above is a $q$-ary 1-perfect code and has a length equal to $(q - 1)nm + n + m$.
\ebtheorem

\proofr
Since $n = (q^{s_1}-1)/(q-1)$ and  $ m = (q^{s_2}-1)/(q-1)$,   then $$(q - 1)nm + n + m =  (q^{s_1 + s_2}-1)/(q-1).$$
Therefore, the length of the code $\mathbb D$ is correct.
Next, we need to show that the number of codewords in the code $\mathbb D$ is correct and the minimum distance
$d(\mathbb D)$ of the code $\mathbb D$ is 3.

Since the $q$-ary  MDS code with parameters $[(q - 1)n + 1, (q - 1)n, 2]_q$ contains  $q^{(q -1) n}$  codewords and each coset, formed by this code, is divided into $(q - 1)n + 1$ subcodes
${\mathbb C}_{j_i}^{r_i}$, then
$$|{\mathbb C}_{j_i}^{r_i}|((q - 1)n + 1) = q^{(q -1) n}.$$
The cardinality  $|{\mathbb C}_{j_0}| = q^{n - s_1}$  for all $j_0 \in \{ 0, 1, \ldots,   (q - 1)n \}$.
Hence, for each ${\bf r} \in \mathbb R$, we can construct
$$|{\mathbb C}_{j_0}||{\mathbb C}_{j_i}^{r_i}|^m((q - 1)n + 1)^m =
q^{n-s_1}q^{(q - 1)nm} = q^{(q - 1)nm + n - s_1}$$
codewords.
Since the cardinality  $|{\mathbb R}| = q^{m - s_2}$, then
$$ |\mathbb D| = q^{(q-1)nm + n + m - (s_1 + s_2)}.$$

Now we show that the minimum distance $d(\mathbb D)$ = 3.
Assume that the vectors  ${\bf x} = ({\bf x}_0|{\bf x}_1|\cdots |{\bf x}_m)$  and
${\bf y} = ({\bf y}_0|{\bf y}_1|\cdots |{\bf y}_m)$  belong to the code $\mathbb D$.
Then $$d({\bf x},{\bf y}) \geq \sum_{i = 0}^m d({\bf x}_i,{\bf y}_i),$$
where the vectors ${\bf x}_0$, ${\bf y}_0$ have length $n$, and for $1 \leq  i \leq m$  the vectors
${\bf x}_i$, ${\bf y}_i$ have length $n + 1$.
Let $r_i = p({\bf x}_i)$ and $r'_i = p({\bf y}_i)$, $i = 1, 2, \ldots, m$.
Then the vectors ${\bf r} = (r_1, r_2, \ldots, r_m)$ and ${\bf r}' = (r'_1, r'_2, \ldots, r'_m)$
belong to $\mathbb R$.

If $d({\bf x}_i,{\bf y}_i) = 0$, then ${\bf r} = {\bf r}'$. If ${\bf r} \neq {\bf r}'$, then
$d({\bf x}_i,{\bf y}_i) \geq 1$. Therefore, if $d({\bf r},{\bf r}') \geq 3$, then $d({\bf x}_i,{\bf y}_i) \geq 1$
for all values of  $i$. Thus,

$$\sum_{i = 0}^m d({\bf x}_i,{\bf y}_i) \geq 3 \, \, \mbox{for}   \, \, {\bf r} \neq {\bf r}'. $$

If ${\bf r} = {\bf r}'$, then $p({\bf x}_i) = p({\bf y}_i)$ and $d({\bf x}_i,{\bf y}_i) \geq 2$
for ${\bf x}_i \neq {\bf y}_i$. Assume that ${\bf x}_0 \in {\mathbb C}_{j_0}$,  ${\bf y}_0 \in {\mathbb C}_{k_0}$, ${\bf x}_i \in {\mathbb C}_{j_i}^{r_i}$ and ${\bf x}_i \in {\mathbb C}_{k_i}^{r_i}$, where $i = 1, 2, \ldots, m$.
Then the equality  $d({\bf x}_i,{\bf y}_i) = 0$ means that $j_i = k_i$,  $i = 0, 1, \ldots, m$;
since ${\bf j} = (j_0, j_1, \ldots, j_m)$  and ${\bf k} = (k_0, k_1, \ldots, k_m)$ can coincide only in $m - 1$
positions, then for at least one value $i \in \{0, 1, \ldots, m\}$, the following inequality holds
$d({\bf x}_i,{\bf y}_i) \geq 2$, and for some other value of $i$, the following inequality holds
$d({\bf x}_i,{\bf y}_i) \geq 1$.
Therefore,  $d({\bf x},{\bf y})\geq 3$ with the exception of the case when ${\bf j} = {\bf k}$.
However, in this case, if ${\bf x}_i \neq {\bf y}_i$,
then $d({\bf x}_i,{\bf y}_i) \geq 3$ and  $d({\bf x},{\bf y})\geq 3$.
\proofend

Next, we present a generalized concatenation construction for $q$-ary codes with parameters
$(q^{s_1 + s_2}, q^{q^{s_1 + s_2} - (s_1 + s_2) - 1}, 3)_q $.
This construction is a $q$-ary version of the well-known generalized concatenation construction for binary extended 1-perfect codes, which was proposed by Phelps in \cite{ph4}.

A vector ${\bf x}  = (x_1, x_2, \ldots, x_n)\in \mathbb{F}_{q}^n$   is \emph{even-like} provided that
$p({\bf x}) = \sum_{i = 1}^n x_i = 0$.
A $q$-ary code is  called \emph{even-like} if it has only even-like codewords. A $q$-ary even-like code of length $n$ is a subcode of the $q$-ary  MDS code with parameters $[n, n - 1, 2]$.

Let $(q - 1)n + 1 = q^{s_1}$. We consider a $q$-ary  MDS code that  has parameters
$[(q - 1)n + 1, n, 2]_q$.  We also  consider the cosets generated by this code. Next, let $ 0  \leq k \leq q - 1$ and let
$ {\mathbb C}_{0}^k, {\mathbb C}_{1}^k, \ldots, {\mathbb C}_{n}^k $ is a partition of the $k$th coset into $q$-ary codes with parameters  $((q -1)n + 1, q^{n - s_1 }, 3)_q $. Such  partitions exist  \cite {rom2}.
Further, let $(q - 1)m + 1 = q^{s_2}$.   We consider  a $q$-ary even-like code $\mathbb R$ with parameters
$((q - 1)m + 1, q^{m - s_2 }, 3)_q $.  Such  codes exist \cite {rom2}.
For each code word ${\bf r} \in \mathbb R$, we define an $(q - 1)m$-ary quasigroup $q_{\bf r}$
of order $(q - 1)n + 1$.
A quasigroup is defined on the index set $\{ 0, 1, \ldots,   (q - 1)n \}$.

Then we form the $q$-ary code ${\mathbb D}$ of length $((q - 1)n + 1)((q - 1)m + 1) =  q^{s_1 + s_2}$  as follows:
\begin{multline*}
{\mathbb D} = \{({\bf c}_0| {\bf c}_1 | {\bf c}_2 | \cdots | {\bf c}_{(q - 1)m} )\, \,  |  \, \,
{\bf c}_i \in {\mathbb C}_{j_i}^{r_i}, {\bf r} = (r_0, r_1, \ldots, r_{(q - 1)m}) \in {\mathbb R}, \\
j_0 =q_{\bf r}(j_1, j_2, \ldots, j_{(q - 1)m}),   j_i \in \{ 0, 1, \ldots,   (q - 1)n \},
i = 0, 1, \ldots, (q - 1)m \}.
\end{multline*}
\bbtheorem
\label{th8}
The  code ${\mathbb D}$ constructed above is a $q$-ary   even-like code and  has parameters
$(q^{s_1 + s_2}, q^{q^{s_1 + s_2} - (s_1 + s_2) - 1}, 3)_q $. For $q = 2$, the minimum distance of the code ${\mathbb D}$ is $4$.
\ebtheorem
\proofr
Since  $(q - 1)n + 1 = q^{s_1}$ and $(q - 1)m + 1 = q^{s_2}$,  then
$$((q - 1)n + 1)((q - 1)m + 1) =  q^{s_1 + s_2}.$$
Hence, the length of the code $\mathbb D$  is  $q^{s_1 + s_2}$.

Next, we need to show that the number of codewords in the code $\mathbb D$ is equal to
$q^{q^{s_1 + s_2} - (s_1 + s_2) - 1}$,  the minimum distance $d(\mathbb D)$ of the code  $\mathbb D$ is 3  and the code  $\mathbb D$  is a $q$-ary   even-like code.

Since the $q$-ary  MDS code with the parameters $[(q - 1)n + 1, (q - 1)n, 2]_q$  contains $q^{(q - 1)n}$ codewords and each  coset formed by this code is divided into $(q - 1)n + 1$ subcodes ${\mathbb C}_{j_i}^{r_i}$, then
$$|{\mathbb C}_{j_i}^{r_i}|((q - 1)n + 1) = q^{(q - 1)n}.$$
Cardinality $|{\mathbb C}_{j_i}^{r_i}| = q^{(q - 1)n - s_1 }$  for all $j_i \in \{ 0, 1, \ldots,   (q - 1)n \}$,
$i = 0, 1, \ldots, (q - 1)m$.
Hence, for each ${\bf r} \in \mathbb R$, we can construct
$$|{\mathbb C}_{j_0}^{r_0}||{\mathbb C}_{j_i}^{r_i}|^{(q - 1)m}((q - 1)n + 1)^{(q - 1)m} =
 q^{(q - 1)n(q - 1)m + (q - 1)n - s_1}$$
codewords.

Since  the cardinality $|{\mathbb R}| = q^{(q - 1)m - s_2 }$, then
$$ |\mathbb D| = q^{(q - 1)n(q - 1)m + (q - 1)n + (q - 1)m + 1 - (s_1 + s_2) - 1} =
q^{((q - 1)n + 1)((q - 1)m + 1) - (s_1 + s_2) - 1}.$$
Consequently,
$$|\mathbb D| = q^{q^{s_1 + s_2} - (s_1 + s_2) - 1}.$$

Now we show that the minimum distance $d(\mathbb D)$ = 3.
Assume that the vectors ${\bf x} = ({\bf x}_0|{\bf x}_1|\cdots |{\bf x}_{(q - 1)m})$ and
${\bf y} = ({\bf y}_0|{\bf y}_1|\cdots |{\bf y}_{(q - 1)m})$ belong to the code  $\mathbb D$.
Then
$$d({\bf x},{\bf y}) \geq \sum_{i = 0}^{(q - 1)m} d({\bf x}_i,{\bf y}_i),$$
where the vectors ${\bf x}_i$, ${\bf y}_i$ have  length   $(q - 1)n + 1$  for  $0 \leq  i \leq (q - 1)m$.
Let $r_i = p({\bf x}_i)$ and $r'_i = p({\bf y}_i)$, $i = 1, 2, \ldots, (q - 1)m$.
Then the vectors ${\bf r} = (r_1, r_2, \ldots, r_{(q - 1)m})$ and ${\bf r}' = (r'_1, r'_2, \ldots, r'_{(q - 1)m})$  belong to the code $\mathbb R$.
If $d({\bf x}_i,{\bf y}_i) = 0$, then ${\bf r} = {\bf r}'$.
If ${\bf r} \neq {\bf r}'$, then $d({\bf x}_i,{\bf y}_i) \geq 1$.
Therefore, if $d({\bf r},{\bf r}') \geq 3$, then $d({\bf x}_i,{\bf y}_i) \geq 1$ for three values of $i$.

Thus,
$$\sum_{i = 0}^{(q - 1)m} d({\bf x}_i,{\bf y}_i) \geq 3 \, \, \mbox{ïðè}   \, \, {\bf r} \neq {\bf r}'. $$

For $q = 2$  we have

$$\sum_{i = 0}^{(q - 1)m} d({\bf x}_i,{\bf y}_i) \geq 4 \, \, \mbox{ïðè}   \, \, {\bf r} \neq {\bf r}'. $$

If ${\bf r} = {\bf r}'$, then $p({\bf x}_i) = p({\bf y}_i)$ and $d({\bf x}_i,{\bf y}_i) \geq 2$
for ${\bf x}_i \neq {\bf y}_i$.
Assume that
${\bf x}_0 \in {\mathbb C}_{j_0}$,  ${\bf y}_0 \in {\mathbb C}_{k_0}$, ${\bf x}_i
\in {\mathbb C}_{j_i}^{r_i}$ and ${\bf x}_i \in {\mathbb C}_{k_i}^{r_i}$, where  $i = 1, 2, \ldots, (q - 1)m$.
Then the equality  $d({\bf x}_i,{\bf y}_i) = 0$ means that $j_i = k_i$,   $i = 0, 1, \ldots, (q - 1)m$;
since ${\bf j} = (j_0, j_1, \ldots, j_{(q - 1)m})$ and ${\bf k} = (k_0, k_1, \ldots, k_{(q - 1)m})$ can coincide only in $(q - 1)m - 1$
positions, then $d({\bf x}_i,{\bf y}_i) \geq 2$  for at least two values of $i$.
Consequently, $d({\bf x},{\bf y})\geq 4$ with the exception of the case when  ${\bf j} = {\bf k}$.
However, in this case, if  ${\bf x}_i \neq {\bf y}_i$,
then $d({\bf x}_i,{\bf y}_i) \geq 3$ and  $d({\bf x},{\bf y})\geq 3$ (respectively,  $ d({\bf x}_i,{\bf y}_i) \geq 4$ and $d({\bf x},{\bf y})\geq 4$ for $q =2$).

It remains to show that the code  $\mathbb D$  is a $q$-ary   even-like code.
Since
$$p({\bf x}) = p(p({\bf x}_0), p({\bf x}_1), \ldots, p({\bf x}_{(q - 1)m}))
= p(r_0, r_1, \ldots, r_{(q - 1)m}) = p(\bf r)$$
and $p(\bf r) = 0$ for all ${\bf r} \in \mathbb R$, then $p(\bf x) = 0$ for all ${\bf x} \in \mathbb D$.
\proofend

For $q = 2, 3$, the MDS codes with the minimum distance 2 are unique.
The above constructions are applicable to any partitions of the space $\mathbb{F}_{q}^{n}$ into MDS codes with  minimum distance  2.

A generalized  concatenation construction can be considered as a generalized direct product of codes.
This construction is also applicable to  MDS codes with  minimum distance  2.
Thus, a generalized concatenation construction allows us to construct not only the code ${\mathbb D}$ with parameters $(q^{s_1 + s_2}, q^{q^{s_1 + s_2} - (s_1 + s_2) - 1}, 3)_q $
but  allows us to also construct a $q$-ary MDS code (of the length $q^{s_1 + s_2}$ with  minimum  distance 2) that contains the code  ${\mathbb D}$.

Assume that the code $\mathbb R$ is an element of some partition of the MDS code into subcodes with parameters
$(q^{s_2}, q^{q^{s_2} - s_2  - 1}, 3)_q $.
Then each element of this partition allows us to construct $q^{s_1}$  disjoint codes with parameters
$(q^{s_1 + s_2}, q^{q^{s_1 + s_2} - (s_1 + s_2) - 1}, 3)_q $.
The partition of the $q$-ary MDS code into subcodes with parameters $(q^{s_2}, q^{q^{s_2} - s_2  - 1}, 3)_q $ contains $q^{s_2}$ elements.

Consequently, if a $q$-ary code $\mathbb R$  is an element of some partition of the $q$-ary MDS
code (of length  $q^{s_2}$ with  minimum distance $2$) into subcodes that have  parameters
$(q^{s_2}, q^{q^{s_2} - s_2  - 1}, 3)_q $, then the code
$\mathbb D$  also is  an element of some partition of the $q$-ary MDS  code (of length $q^{s_1 + s_2}$ with minimum distance $2$) into subcodes that have parameters
$(q^{s_1 + s_2}, q^{q^{s_1 + s_2} - (s_1 + s_2) - 1}, 3)_q $.

 \end{document}